\newif{\ifjournal}
  \journaltrue

\ifjournal
  \documentclass[]{aa}
  \usepackage{graphicx,txfonts,amssymb,natbib,subfigure}
  \authorrunning{M. Maturi et al.}
  \titlerunning{Galaxy cluster merger kinematics by Rees-Sciama effect}
\else
  \documentclass{paper}
\fi

\begin{document}

\title{Galaxy cluster merger kinematics by Rees-Sciama effect}

\ifjournal
  \author{Matteo Maturi\inst{1,2,3,4},
    Torsten En{\ss}lin\inst{4},
    Carlos Hern\'andez--Monteagudo\inst{4,5},
    Jos\'e Alberto Rubi\~no--Mart\'in\inst{4,6}}
    \institute{
    $^1$ Dipartimento di Astronomia, Universit\`a di Padova, Vicolo
    dell'Osservatorio 2, 35120 Padova, Italy \\
    $^2$ ITA, Universit\"at Heidelberg,
    Albert-\"Uberle-Str.~2, 69120 Heidelberg, Germany\\
    $^3$ Dipartimento di Astronomia, Universit\`a di Bologna, Via
    Ranzani 1, 40127 Bologna, Italy\\
    $^4$ Max-Planck-Institut f\"ur Astrophysik, P.O. Box 1523,
    85740 Garching, Germany\\
    $^5$ Dept. of Physics \& Astronomy, University of Pennsylvania, 
    209 South 33rd Str. Philadelphia, PA 19104-6396 USA\\
    $^6$ Instituto de Astrof\'isica de Canarias, C/V\'ia L\'actea s/n, 
    E-38200 Tenerife, Spain}
\else
  \author{Matteo Maturi\inst{1,2,3,4},
    Torsten En{\ss}lin\inst{4},
    Carlos Hern\'andez--Monteagudo\inst{4,5},
    Jos\'e Alberto Rubi\~no--Mart\'in\inst{4,6}}
    \institute{
    $^1$ Dipartimento di Astronomia, Universit\`a di Padova, Vicolo
    dell'Osservatorio 2, 35120 Padova, Italy \\
    $^2$ ITA, Universit\"at Heidelberg,
    Albert-\"Uberle-Str.~2, 69120 Heidelberg, Germany\\
    $^3$ Dipartimento di Astronomia, Universit\`a di Bologna, Via
    Ranzani 1, 40127 Bologna, Italy\\
    $^4$ Max-Planck-Institut f\"ur Astrophysik, P.O. Box 1523,
    85740 Garching, Germany\\
    $^5$ Dept. of Physics \& Astronomy, University of Pennsylvania, 
    209 South 33rd Str. Philadelphia, PA 19104-6396 USA\\
    $^6$ Instituto de Astrof\'isica de Canarias, C/V\'ia L\'actea s/n, 
    E-38200 Tenerife, Spain}
\fi

\date{\emph{Astronomy \& Astrophysics, submitted}}

\def\vk{\vec{k}}
\def\vr{\vec{r}}
\def\vrp{\vec{r}_\perp}
\def\vp{\vec{v}_\perp}
\def\vZP{\vec{Z}_\Psi}
\def\vZ{\vec{Z}}
\def\vV{\vec{V}}
\def\be{\begin{equation}}
\def\ee{\end{equation}}

\newcommand{\abstext}{We discuss how to use the Rees-Sciama (RS) effect
  associated with merging clusters of galaxies to measure their kinematic
  properties. In a previous work (Rubi\~no-Mart\'in et al. 2004), the
  morphology and symmetries of the effect were examined by means of a
  simplified model. Here, we use realistic N-body simulations to better
  describe the effect, and to confirm that the signal has a characteristic
  quadrupole structure. From the amplitude of the signal obtained, we conclude
  that it is necessary to combine several cluster mergers in order to achieve
  a detection.
  Using the extended Press-Schechter formalism, we characterized the expected
  distribution of the parameters describing the mergers, and we used these
  results to generate realistic mock catalogues of cluster mergers. To
  optimize the extraction of the RS signal, we developed an extension of the
  spatial filtering method described in \cite{HN96.1}. This extended filter
  has a general definition, so it can be applied in many other fields, such as
  gravitational lensing of the CMB or lensing of background galaxies. It has
  been applied to our mock catalogues, and we show that with the announced
  sensitivities of future experiments like the Atacama Cosmology Telescope
  (ACT), the South Pole Telescope (SPT) or the Atacama Large Millimeter Array
  (ALMA), a detection of the signal will be possible if we consider of the
  order of $1,000$ cluster mergers.}

\abstract{\abstext} 
 
\maketitle 

\section{Introduction}

Observations of the fluctuations of the cosmic microwave background (CMB) can
give us information about the formation of structures in the Universe, because
the evolution of the gravitational potentials leaves its imprint on the CMB
photons as they travel along them.  This physical effect is usually split into
two terms.  One is the integrated Sachs-Wolfe effect
\citep[ISW;][]{SA67.1,HU94.1}, which is produced by the linear evolution of
the potentials. The other is the Rees-Sciama effect \citep[RS;][]{RE68.1,
MA90.2, SE96.2}, in which the density contrast producing the gravitational
potential is in its non-linear regime.

In this work we discuss the RS effect associated with the non-linear regime of
galaxy clusters mergers. In a previous work (\cite{RU04.1}, hereafter Paper
I), this problem was examined using a simplified model of the physics of the
merger. The amplitude, morphology and symmetries of the effect were discussed
and characterized in terms of the physical parameters of the merger. However,
the RS effect is {\em not} the only signal being generated by clusters. The
presence of hot gas in the Intra--Cluster Medium (ICM) and the peculiar
velocity of the cluster with respect to the Hubble flow introduce temperature
fluctuations which are proportional to the integral of pressure (thermal
Sunyaev-Zel'dovich effect, [tSZ]) and the momentum density (kinematic
Sunyaev-Zel'dovich effect, [kSZ]) along the line of sight,
\citep{SU72.1,SU80.1}. In addition, galaxy clusters may host radio-galaxies or
infrared galaxies whose flux might be relevant to the small-amplitude effects
we are studying here. Further, the intrinsic CMB field and its deflection
caused by the matter distribution in the cluster produce different
anisotropies which must be accounted for.  In Paper I, it was proposed to
co-add coherently the RS signals from a sample of cluster mergers in order to
achieve a statistical detection of the effect against these ``foreground''
anisotrpy signals.

Here, we extend this work in three ways. First, we use hydrodynamical
simulations in order to compute realistic maps of the RS effect, and to show
that the morphology of the signal has the characteristic quadrupole structure
described in Paper I. We also show that N-body simulations contain an
additional signal which originates during the collapse of the environmental
dark-matter in the potential well of the system, although the morphology of
this signal is completely different from the one associated with the merging
induced one.
Second, we perform a detailed study of the expected distribution of the
parameters describing a merger of galaxy clusters (masses of the components,
distances and velocities), and we present semianalytical expressions for the
probability of finding a cluster merger with a given parameter set. These
equations can be used to produce a realistic ``catalogue'' of cluster mergers,
which we use in order to forecast the detectability of the effect.
Finally, we investigated the prospects for the RS signal extraction from
cluster mergers by future CMB experiments, like the South Pole
Telescope\footnote{SPT, see {\texttt http://spt.uchicago.edu/}.} or the
Atacama Cosmology Telescope\footnote{ACT, see {\texttt
http://www.hep.upenn.edu/$\sim$angelica/act/act.html}.}. To this end, we
extended the filtering method described in Haehnelt \& Tegmark (1996). Our new
filter has a general mathematical definition, so it could be applied to
extract the signal in many other fields, e.g. gravitational lensing of the CMB
\citep{SE96.2,MAT04.1} or lensing of background galaxies
\citep[e.g.][]{MAT04.2}. We use this extended filter to separate the RS signal
from the following three components: the primordial CMB, the gravitational
lensing imprint, and the kinetic SZ effect. We show that with upcoming
instruments, the RS signal could be detected if we consider of the order of
1,000 observed cluster mergers.  As we will see below, the main limiting
``noise'' component in such a measurement will be the lensed CMB.

\section{The Rees-Sciama effect} \label{sec:RS}

In Paper I, the Rees-Sciama effect associated with mergers of galaxy clusters
was discussed using an analytical model for the merger. In addition, an
analytical prescription to prepare Rees-Sciama maps from hydrodynamical
simulations was introduced.  Here, we implement this prescription and we
present some RS maps resulting from numerical simulations. The morphology of
the effect can then be directly compared to that predicted in Paper I.

In order to prepare a map of the RS effect from a hydrodynamic simulation, we
use the analytic recipe proposed in Paper I \citep{RU04.1} to compute the CMB
temperature change along the line-of-sight (LOS) direction $\hat{n}$,
\be
\label{eq:RS}
\frac{\delta T_{\rm RS}}{T} (\vrp) = -\frac{4\,G}{c^3}\, 
\int d^3 r' \varrho(\vr')\, \vec{v}(\vr') \vec{\cdot} \frac{\vrp -
\vrp'}{|\vrp - \vrp'|^2},
\ee
where $\varrho(\vr )$ is the total mass density and $\vec{v}(\vr)$ the mean
velocity of matter. In this equation, the vectors are split into line-of-sight
(LOS) parallel and LOS-perpendicular components, so $\vr = (\vr_\perp,
z)$. The axis along the line-of-sight is specified by $\hat{n}$, with the
positive values pointing towards the observer. To be valid, this expression
requires the gravitational fields to be weak (which is the case for galaxy
clusters), and the matter distribution to be nearly stationary (i.e. $v/c \ll
1$), so the mass density does not show an explicit dependence on time. The
algorithm was integrated into the SMAC code\footnote{(Simulated MAps Creator:
http://dipastro.pd.astro.it/~cosmo)}, which aims at deriving maps for the main
galaxy cluster observables from numerical simulations.

Due to its gravitational nature, the RS effect can also be related to the
deflection angle, yielding an expression very useful for analytical
computations:
\be
\label{eqn:RS2}
\frac{\delta T_{\rm RS}}{T} (\vrp) = -\frac{v}{c}\, \sin \beta 
\, \vec{u}_{v\, \perp} \cdot \vec{\alpha} (\vec{n}) \, .
\ee
Here $\vec{\alpha}$ is the gravitational lensing deflection field of the
cluster, $\vec{u}_{v\, \perp}$ is the unitary transverse velocity vector and
$\beta$ is the angle described by the cluster velocity $\vec{v}$ and the plane
of the sky, as shown by \cite{BI83.1}.

The dot product $\vec{u}_{v\, \perp} \cdot \vec{\alpha} (\vec{n})$ has a
dipolar structure elongated along $\vec{u}_{v\, \perp}$ and, according to
Equation~(\ref{eqn:RS2}), the RS anisotropy is proportional to the halo mass
and to its transverse velocity. Due to these dependencies, forming galaxy
clusters with two major merging haloes seem to be the best candidates for RS
detection. First, because of the large masses and velocities involved, of the
order of $10^{14} \, M_\odot$ and $1,000 \, {\rm km/s}$ respectively. Second,
because the contributions of the merging components add constructively due to
the converging cluster motions.

A simple description of this scenario can be built by considering two dark
matter haloes, having a NFW profile \citep{NA95.3}, in free fall. It provides
a simple derivation for the lensing and the RS effects, from which it is
straightforward to obtain the templates necessary for the data reduction
(Section~\ref{sec:strategy}). The RS effect can be computed according to
Equation~(\ref{eqn:RS2}) where the deflection angle can be computed
analytically \citep{BA96.1}. As it will be shown, the fact that the deflection
field is axially symmetric on large scales justifies the assumption of
the halos being spherical.

\begin{figure}[!ht]
\begin{center}
  \includegraphics[width=4.2truecm]{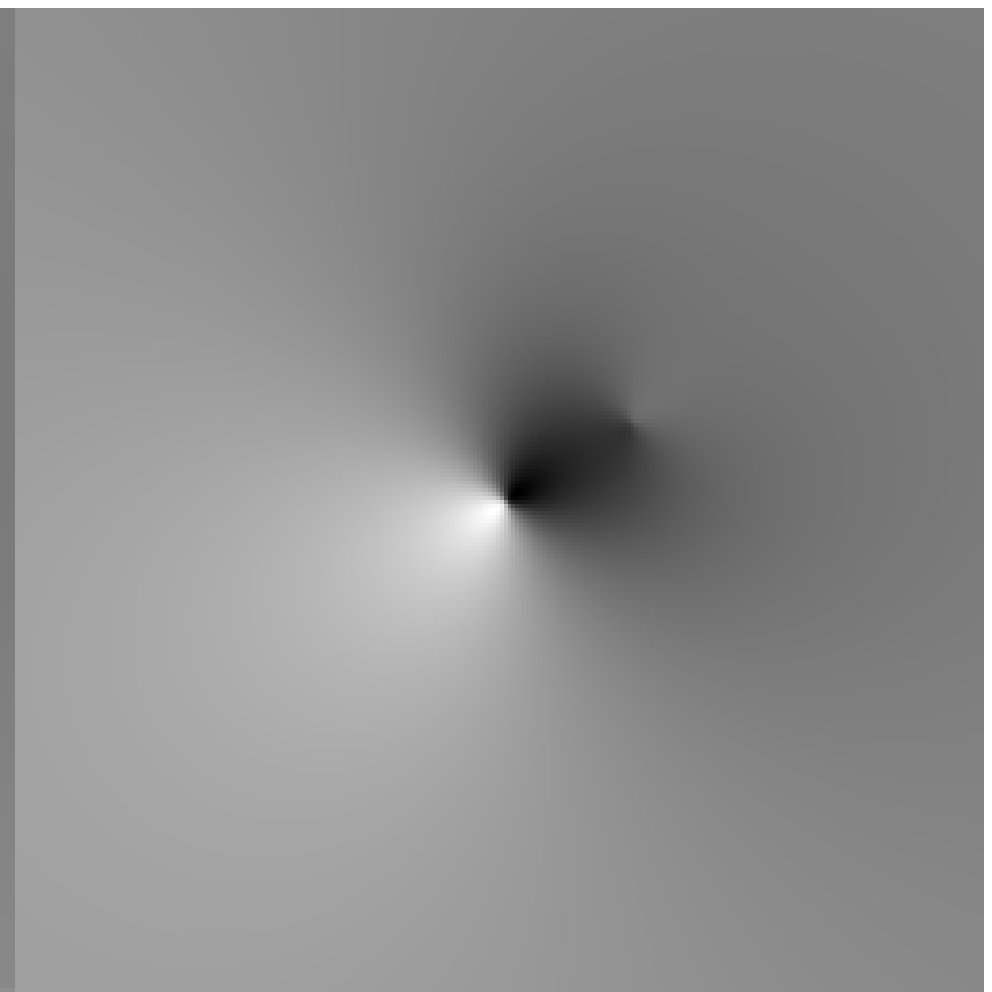}
  \includegraphics[width=4.2truecm]{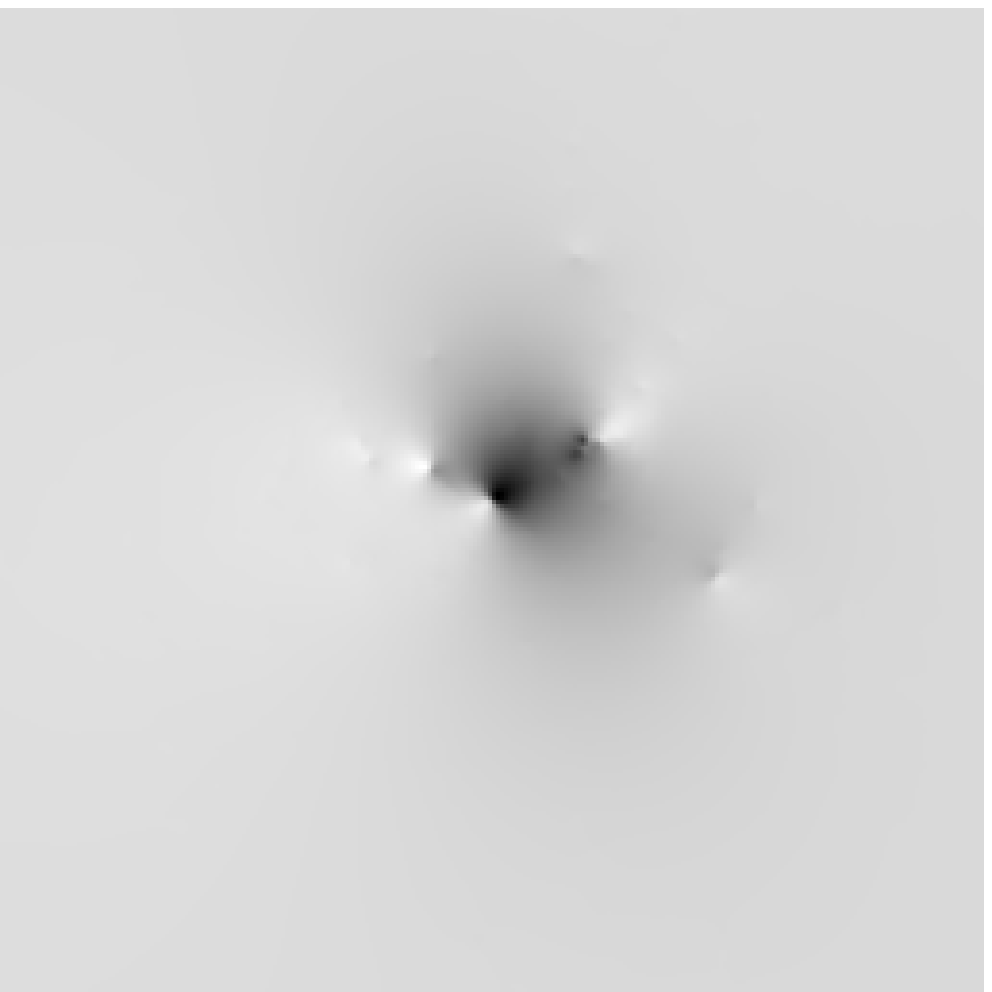}
  \caption{A comparison between our analytical description of the RS signal
    generated in a major merger (left panel) and the numerical simulation
    g24+200 described in the text and in \cite{RA04.1} (right panel). The
    field of view is $2$ degrees on a side. The two clusters are placed at
    redshift $0.3$, have masses of $M_1=6\times 10^{14} M_\odot/h$ and
    $M_2=10^{15} M_\odot/h$ respectively, and they are falling onto each other
    with a relative velocity of $1800$ km/s. The analytical model represents
    the same merger of the numerical simulation, showing that our simple
    recipe provides a good description of the cluster merger. The total
    contribution of minor mergers is negligible, as explained in
    Section~(\ref{sec:filter}). The main difference is that the simulation
    includes the roughly spherical collapse of environmental DM onto the
    potential well. This collapse-induced signal has the shape of a negative
    peak with circular symmetry centered on the potential minimum. This
    contribution adds coherently to the signal of the two main halo mergers
    and is strictly related to the merging dynamic we want to estimate. Since
    we are ignoring this term, the resulting signal amplitude is smaller,
    making this choice conservative. In any case our filter can easily include
    this component, providing an estimate of the dynamic of the whole
    infalling dark matter and not only of the main merger components.}
  \label{fig:RS_numeric2}
\end{center}
\end{figure}

As an example, we present in Figure~(\ref{fig:RS_numeric2}) the RS effect of a
major merger event. In the left panel, the RS signal was computed with our
simple analytical recipe, whereas in the right panel we used the numerically
simulated cluster g24+200 described in \cite{RA04.1}. This cluster consists of
two merging haloes, of masses $M_1=6\times 10^{14} M_\odot/h$ and $M_2=10^{15}
M_\odot/h$, with an infall velocity of $1800\,km/s$, provided by
G. Tormen. The field of view is $2$ degrees on a side.  This figure
demonstrates that our simple analytical model can describe the RS signal
generated in major cluster mergers with reasonable accuracy. The main
differences between our description and the output obtained from the numerical
simulations are related to the presence of minor merging substructures within
the main object, and the roughly spherical collapse of environmental dark
matter (DM) onto the potential.

The effect of minor mergers can be neglected, since their amplitude is small
and their contribution vanishes after applying our filter (see
Section~\ref{sec:filter} below).  The roughly spherical collapse of
environmental DM has the shape of a negative circular symmetric peak centered
on the potential minimum, and adds coherently to the merger-induced RS
signal. Also this extra RS component is related to the dynamical status of DM
and could be considered part of the signal we are searching for. The inclusion
of this contribution in our procedure can be easily implemented by simply
changing the filter RS template (see Section~\ref{sec:filter}). In this case
we would measure the full DM momentum instead of only the DM momentum of the
main mergers. Since we ignore this signal, our estimate of the detectability
of the RS signal is conservative.

\section{Distribution of physical parameters describing the merger}\label{sec:carlos}

We investigate the probability distribution of the physical parameters
describing a merger.  We use the approach of \citet{LA93.1} based on the
extended Press-Schechter formalism to model the distribution of cluster
mergers in the Universe.  This description basically relies on the probability
of a region of mass $M_1$ having overdensity $\delta_1$ {\em provided} that,
at the same point, the overdensity measured on a larger scale (corresponding
to a mass $M_2$) is $\delta_2$. In other words, our starting point is the
conditional probability of having $\delta_1$ on scales of $M_1$, {\em given}
that on scales corresponding to $M_2$ the density contrast is $\delta_2$:
\begin{equation}\label{eq:ps1}
  \begin{array}{ll}
    \displaystyle
    f(\delta_1, M_1 \big| \delta_2, M_2) dS_1 = &
    \displaystyle
    \frac{\delta_1 - \delta_2}{\sqrt{2\pi}(S_1 - S_2)^{3/2}} \times\\
    &
    \displaystyle
    \exp{\biggl[ -\frac{(\delta_1 - \delta_2)^2}{2(S_1-S_2)} \biggr]} dS_1\,.
  \end{array}
\end{equation}
Here, $S_i$ refers to the variance of the $\delta$ field on the scale of mass
$M_i$, $S_i \equiv \langle \sigma (M_i)^2 \rangle$. Via the Bayes theorem, it
is possible to compute the reverse probability, i.e., the probability of
having a region of mass $M_2$ with overdensity $\delta_2$ given that on scale
of $M_1$ the density contrast is $\delta_1$. If one introduces now the rate of
growth of density fluctuations in linear theory, then this reverse probability
can be understood as the rate at which a given halo collects matter:
\begin{equation}\label{eq:nm}
  \begin{array}{lll}
    \displaystyle
    {\cal R}(M_1,\Delta M, z) \equiv & \frac{d^2p}{d\Delta M dz} = &
    \displaystyle
    \sqrt{\frac{2}{\pi}} \frac{\delta_{sc}}{D(z)^2S_2} \times \\
    & &
    \displaystyle
    \biggl| \frac{dD(z)}{dz} \biggr|
    \biggl| \frac{d\sigma (M_2)}{dM_2}\biggr|
    \frac{1}{\left(1-\frac{S_2}{S_1}\right)^{3/2}} \times\\
    & &
    \displaystyle
    \exp{\left[ -\frac{\delta_{sc}^2}{2D(z)^2} \;
	\left(\frac{1}{S_2} - \frac{1}{S_1} \right)\right]  }\,,
  \end{array}
\end{equation}
where, $\delta_{sc}$ stands for the spherical collapse critical overdensity,
and $\Delta M$ is the amount of mass being collected by the halo of mass $M_1$,
so that $M_2 = M_1 + \Delta M$.  The quantity $D(z)$ is the linear growth
factor of density perturbations.  When compared to numerical simulations, this
formalism is able to reproduce approximately the halo merger history
\citep[e.g.][]{VA02.1}.  However, as noted by \citet{BE05.1}, it has some
intrinsic inconsistency due to its asymmetry in the arguments $M_1$ and $\Delta
M$. In fact, the amount of clusters of mass $M_1$ to which clusters of mass
$\Delta M$ are merging does {\em not} equal the number of clusters of mass
$\Delta M$ to which clusters of mass $M_1$ are merging.  We overcome this
problem by defining the number of mergers of clusters of masses $M_1$ and
$\Delta M$ as:
\begin{equation}\label{eq:nmergers}
  \begin{array}{ll}
    \displaystyle
    N_{merg} (M_1, \Delta M, z) \equiv &
    \frac{1}{2} \biggl[{\cal R}(M_1,\Delta M, z) \; n(M_1,z) +\\ 
    &
    \displaystyle
    {\cal R}(\Delta M,M_1, z) n(\Delta M, z) \biggr] \Delta z \,,\\
  \end{array}
\end{equation}
where $n(M,z)$ represents the number density of haloes of mass $M$ at redshift
$z$, \citep{PR74.1}.  Since Equation~(\ref{eq:nm}) provides a merger {\em
rate}, we must multiply it by the finite redshift interval $\Delta z$ which is
taken to be equal to the dynamical time of the final cluster (approximated by
$t_{dyn}(z) \simeq 0.09 / H(z)$, with $H(z)$ the Hubble constant
\citep{TO02.1}.  Although this description of the halo merger rate might not
be accurate, it should still provide a consistent description of the halo
formation history.\\

We follow the model given in, \cite{SA02.2}, when assigning distances and
velocities to the merging clusters. The typical initial distance between a
pair of clusters, following Equation~(10) in \citet{SA02.2}, may be
approximated as
\begin{equation}
  d_0 \simeq 4.5 \;\;\biggl(\frac{t(z)}{10^{10}\,yr} \biggr)^{2/3} \mbox{Mpc},
  \label{eq:d0}
\end{equation}
where $t(z)$ is the cosmic time. We assume that the probability of finding two
merging clusters at a given distance $r\,(0<r<d_0)$ is directly proportional
to the time spent by those clusters in the interval $[r,r+dr]$, that is,
\begin{equation}
  p(r) \propto \frac{1}{v(r)},
  \label{eq:pr}
\end{equation}
where $v(r)$ is the relative infall velocity of the clusters, given by
\begin{equation}
  v(r) \simeq v_0 \;\;\sqrt{\frac{M_1 + \Delta M}{10^{15}\;M_{\odot}}} 
                 \sqrt{\frac{r}{d_1}\biggl(1-\frac{r}{d_0} \biggr)}.
\label{eq:v(r)}
\end{equation}
In this equation, $v_0 = 2930$ km s$^{-1}$ and $d_1=$1 Mpc, and we ignore the
angular momentum of the merging system. Putting all this together, we end up
with a semi-analytical model of the merging cluster population in our
universe, and of their velocities. Finally, we assume an isotropic
distribution for the orientation of the merger plane with respect to the
observer.

\section{Observational strategy}\label{sec:strategy}

The millimetric observations of upcoming experiments will provide extensive
catalogues of galaxy clusters detected through their thermal SZ effect (tSZ),
together with estimates of their masses and their geometrical properties. The
tSZ effect can be measured separately from the other effects due to its unique spectral
dependence. The data of CMB temperature fluctuations will contain signatures
of the RS effect we aim to detect, together with the following components
which will be regarded here as {\it contaminants}: the primordial CMB field,
the instrumental noise, the lensing of the CMB anisotropies induced by the
matter present along the line of sight, and the kinematic SZ (kSZ) effect
generated by the cluster peculiar velocity with respect to the Hubble flow.
As long as the contaminants show different spatial characteristics compared to
the signal we are trying to unveil (the RS effect), it is possible to define a
filter which {\em optimally} reduces the impact of those sources of
noise. Regarding this, instrumental noise will be assumed to be scale--free
white noise, i.e., to have the same power at all scales. The intrinsic CMB
fluctuations usually have (if clusters are not too close) scales larger than
those of the RS signal. On the other hand, the kSZ and the tSZ residuals are
confined to the region of the cluster where the gas is located. Compared to
these scales, the RS is usually broader.

However, if a contaminant shows spatial power on scales which are too close to
those of the signal, only a more direct approach can be performed. This turns
out to be the case for the anisotropies introduced by the lensing of the
intrinsic CMB fluctuations.  This component will require a separate processing
(Section~\ref{sec:de-lens}): from a template of the lensing-induced deflection
field, we shall ``{\it deflect back}'' the data, in an attempt to subtract the
effect of gravitational lensing on the CMB. This will allow us to minimize the
impact of this contaminant.

Once the cluster lensing signal is reduced by this ``de-lensing procedure'',
an optimized filter will be applied to suppress the other noise terms and to
estimate the RS effect amplitude. This amplitude depends on the cluster infall
velocity, which is essentially the physical parameter we are measuring (see
Equation~\ref{eqn:RS2}). This filter incorporates in its definition two
templates: one for the RS effect described in Section~(\ref{sec:RS}),
parameterized by the mass which we assumed to be estimated from observations
of the tSZ effect, and one for the kSZ effect. The second one could be directly
derived from the millimetric observations by observing at $217 \mbox{GHz}$ and
filtering away the CMB features through a high--pass filter.

As we shall see, the resulting signal-to-noise ratio of a single merger is too
small to provide a detectable signal. Thus it is necessary to average multiple
measurements of many mergers. The detection of the cluster through their tSZ
effect is already well described in other papers \citep[e.g.][]{SA04.1} so it
will not be discussed here. We will now describe in detail the steps of the
observational strategy summarized above.

\subsection{De-lensing procedure}\label{sec:de-lens}

The CMB anisotropy produced by the gravitational lensing effect of galaxy
clusters has a typical amplitude ten times larger than the RS anisotropy
induced by the same clusters. These two effects arise from the same
gravitational potential, so that their typical spectral scales are comparable
and correlated in position and, although their patterns are not
identical\footnote{The difference in their patterns arises from the fact that
   the RS effect orientation depends on the velocity direction of clusters,
   while the gravitational lensing anisotropy depends on the local CMB
   temperature gradient direction (see Section~\ref{sec:RS}).},
the similarity of their power spectra complicates the use of any Wiener
filter.

Therefore it is necessary to reduce the lensing effect before the application
of any optimized filter. To achieve this, we propose to remap the observed
data by applying a distortion which compensates, as accurately as possible,
the cluster lensing effect.

To do this it is necessary to invert the lens equation
$T(\vec{\theta})=\tilde{T} (\vec{\theta}-\vec{\alpha}(\vec{\theta}))$ by
introducing the new coordinate
$\vec{\theta}'=\vec{\theta}-\vec{\alpha}(\vec{\theta})$. In this way it is
possible to derive from the observed data $T$ the de-lensed map $\tilde{T}$
thanks to the deflection field template $\vec{\alpha}$, according to
\be\label{eqn:de-lens}
\tilde{T}(\vec{\theta})=T (\vec{\theta}+\vec{\alpha}(\vec{\theta}')) \,.
\ee
Note that the deflection field has to be computed via an iterative procedure,
because it is evaluated at the new coordinate $\vec{\theta}'$.

In order to build a deflection field template, we define the merger using
two NFW haloes according to Equation~(\ref{eqn:RS2}). We assume that their
masses are derived by the tSZ measurements, and for this reason we will
consider some uncertainty in the mass determination.  The assumption of
spherical haloes is justified since the deflection field is roughly symmetric
on large scales. Furthemore, the final estimate of the signal will be produced
after stacking all available merger events.

Of course this procedure will -- apart from the CMB -- also distort the RS
signal, the kSZ signal and the distribution of the instrumental noise.
Because of this, the RS effect and the kSZ templates used in the optimal filter
definition have to be corrected through the same process. This procedure leaves
some randomly distributed residuals in the field, and some non-Gaussian
features in the instrumental noise which will be ignored in the following
analysis.

\subsection{Filter}\label{sec:filter}

Once the cluster lensing contribution is minimized, the remaining data
contaminants in the maps are the CMB primary anisotropy, the kSZ effect and
the instrumental noise.

The Gaussian contribution of CMB and instrumental noise are fully described by
their corresponding power spectra and can be filtered out by optimized
filters.  In particular, the CMB power spectrum shows an exponential cutoff
around the arcminute scale due to Silk damping of primordial
fluctuations. This angular scale is close to the typical one for distant
galaxy clusters, and hence the CMB intrinsic fluctuations should make small
contributions at the spatial frequencies we are interested in. For CMB
experiments with very high--sensitivity on small--scale, such as ACT, SPT or
the Atacama Large Millimeter
Array\footnote{ALMA, see www.eso.org/projects/alma}
we shall assume that the noise is spatially white, i.e., that it introduces
the same amount of power in all scale.

\begin{table}
  \caption{Main characteristics of the upcoming millimetric observatories
  ACT, SPT
  and of our simulated instrument. The second column gives the
  available frequency channels, the third column the angular resolution and
  the last column the relative sensitivity. We are reporting nominal
  sensitivities for ACT, and those of SPT correspond to one hour of
  observation. 
  \label{tab:instruments}}
  \begin{center}
    \begin{tabular}{c|cccc}
      & bands & FWHM & $\Delta$T/beam \\
      & (GHz) & (')& ($\mu K$) \\
      \hline
      ACT  & 145, 225, 265 & 1.7, 1.1, 0.93 & 2, 3.3, 4.7\\
      SPT  &  150, 219, 274 & 1, 0.69, 0.56 & 2.5, 3.0, 2.65\\
      simulation & 217 & 1 & 1 \\
      \hline
    \end{tabular}
  \end{center}
\end{table}

\begin{figure}[t]
  \centering
  \includegraphics[width=7.0truecm]{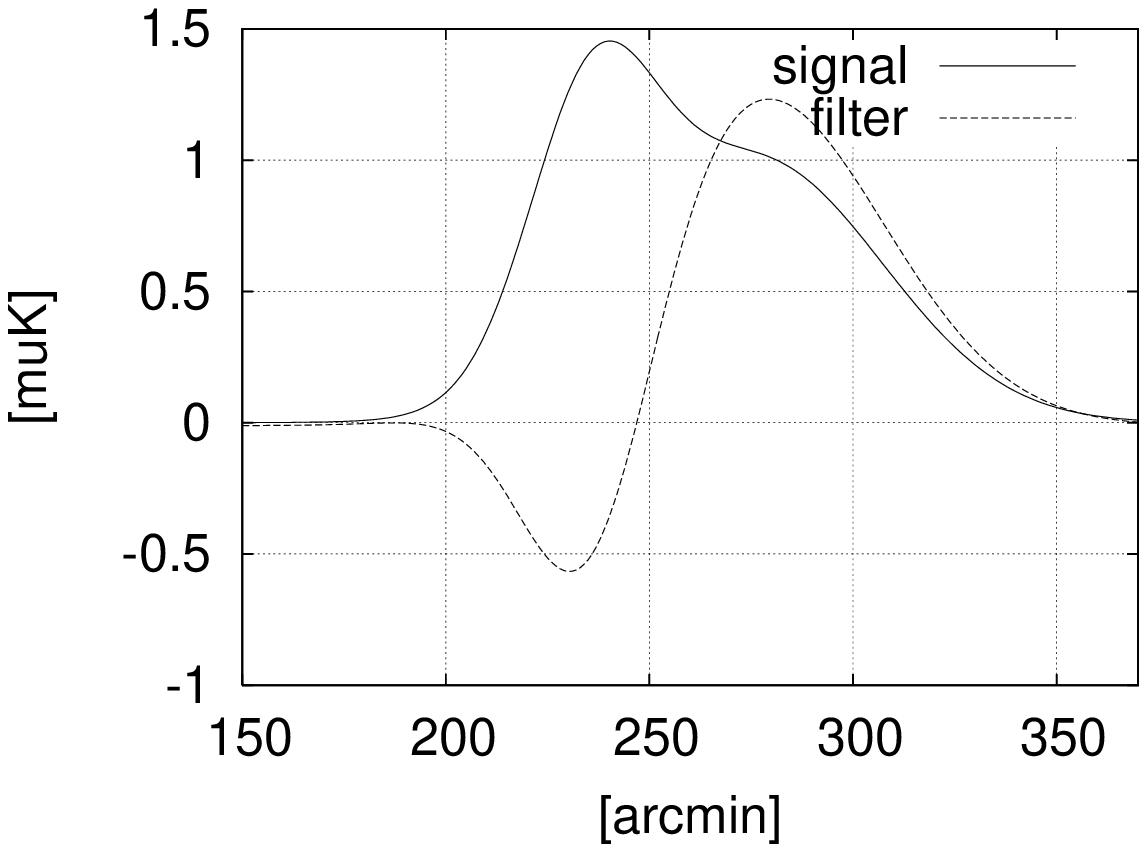}
  \includegraphics[width=7.0truecm]{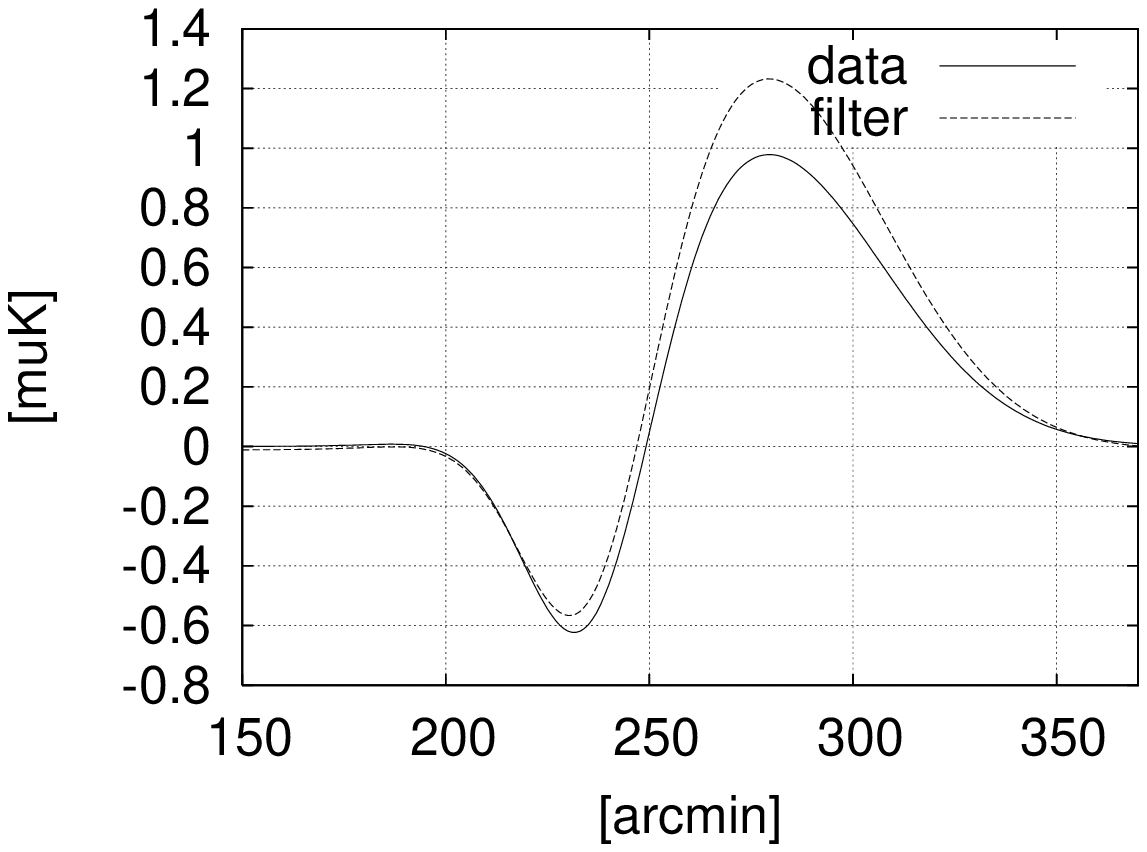}
  \caption{To give an intuitive view of how our filter works, we present here
      two simple examples where the signal and a contaminant of known spatial
      pattern have both Gaussian profiles with the same width. In the upper
      panel both signal and noise are positive and the sum of these
      contributions, the data, is given by the solid line, while the dashed
      line displays the resulting filter. In the bottom panel we only changed
      the sign of the amplitude of the noise, which now is negative. The
      estimator defined in Equation~(\ref{eqn:A}) is the integral of the solid
      line times the dashed one. In the first (upper panel) case, the filter is
      negative where the contaminant is positive, so that their product is
      negative and the contaminant contribution is subtracted. In the second
      case (bottom panel), both filter and contaminant are negative so that
      their product is positive and the negative contribution of the noise is
      compensated.  This actually shows that the same filter provides an
      unbiased estimator also if only the variance of the contaminant
      amplitude is known. That is, the sign of the contaminant is irrelevant
      to the filter, and this reflects the statistical nature of our filter.}
\end{figure}

We present here an optimally matched filter (derived detail in
Appendix~\ref{app:filter}), which maximizes the signal-to-noise ratio by
processing the signal in the Fourier domain. A similar filter construction was
proposed by \cite{HN96.1} in order to extract the kSZ signal from single
clusters. However, that method considered the CMB and instrumental noise the
only sources of confusion when trying to recover the kSZ signal.  Our filter
is defined in a slightly more general context. It provides an estimate of the
amplitude of a signal with a known spatial template and is to be measured
despite a presence of homogeneous noise with a known spatial power spectrum
and an arbitrary number of other contaminants following known spatial
patterns. Hence, the total observed data can be modeled by
\be
s(\vr)=A\tau(\vr) + n(\vr) + \sum_{i=1}^M v_i z_i(\vr) \,,
\ee
where $A$ is the amplitude of the signal, $\tau(\vr)$ is the signal shape,
$n(\vr)$ is a Gaussian random noise with a known power spectrum $P_n(k)$, and
$z_i(\vr)$ are $M$ noise sources with an amplitude $v_i$.  Note that this
model can {\it a priori} be applied to different problems. For instance, in
weak lensing observations, where it is necessary to take into account the
noise introduced by stars, bright galaxies and the cluster core itself, this
filter allows improved results obtained with standard matched filters
\citep[e.g.][]{MAT04.2}.

The Gaussian noise contribution $n(\vr)$ will contain the CMB temperature
fluctuations, the instrumental noise, and all the noise sources which can be
modeled as Gaussian random fields. In this case, the filter derived in
Appendix~(\ref{app:filter}) takes the form
\be \label{eqn:filterRS}
  \Psi(\vk) = \frac{\lambda T(\vk) - B_1 Z_1(\vk) - B_2 Z_2(\vk)}
                   {P_n(k)} \,,
\ee
where $P_n(k)=P_{CMB}(k)+P_{inst.}(k)$ is the total Gaussian noise power
spectrum, including the CMB and the instrumental noise. In our case, we
introduce two spatially characterised contaminants ($M=2$), that is, the kSZ
signal of the two merging haloes $Z_1(\vk)$ and $Z_2(\vk)$.  The normalization
of the first term in the numerator can be expressed as $\lambda=(1+D)/C$, with
\be
  C = \left\{|T|^2/P_n\right\}
\ee
and
\be
  D = \Re \left(\left\{\frac{T^* \left( B_1 Z_1 + B_2 Z_2 \right)}
                            {P_n}\right\}\right) \,.
\ee
The coefficients $B_i$ are given by
\be\label{eqn:B_i}
  B_i = \sigma_{vi}^2 \{ Z_i^* \Psi \} \,,
\ee
and the estimate of the RS effect amplitude $\bar{A}$ will be provided by
\be \label{eqn:A}
  \bar{A} = \left\{ S \Psi^* \right\} \,.
\ee
In these equations, the bracket notation $\{ \,f\,\}$ stands for a double
integral in the k--space of a general function $f(\vk)$ as defined in
Equation~(\ref{eqn:integrand}). The quantity $\sigma_{vi}^2$ represents the
variance of the {\it ith} contaminant amplitude, and the symbol $\Re$ denotes
``real part''.

It is clear from Equation~(\ref{eqn:filterRS}) that the filter can be split
into two components. The first, ($T^*/P_n$), maximizes the sensitivity at the
spatial frequencies where the signal $T$ is large and the noise power spectrum
$P_n$ is small, as opposed to the second ($ -(B_1 Z_1 + B_2 Z_2)/P_n$), which
underweights the regions where the $Z_i$ components are important and
correlated with the filter itself. Note that since the filter $\Psi$ is
present in the definition of the $B_i$ coefficients, the final form of the
filter must be computed iteratively.

\begin{figure}[t]
  \centering
  \subfigure
      {\includegraphics[width=4.2truecm]{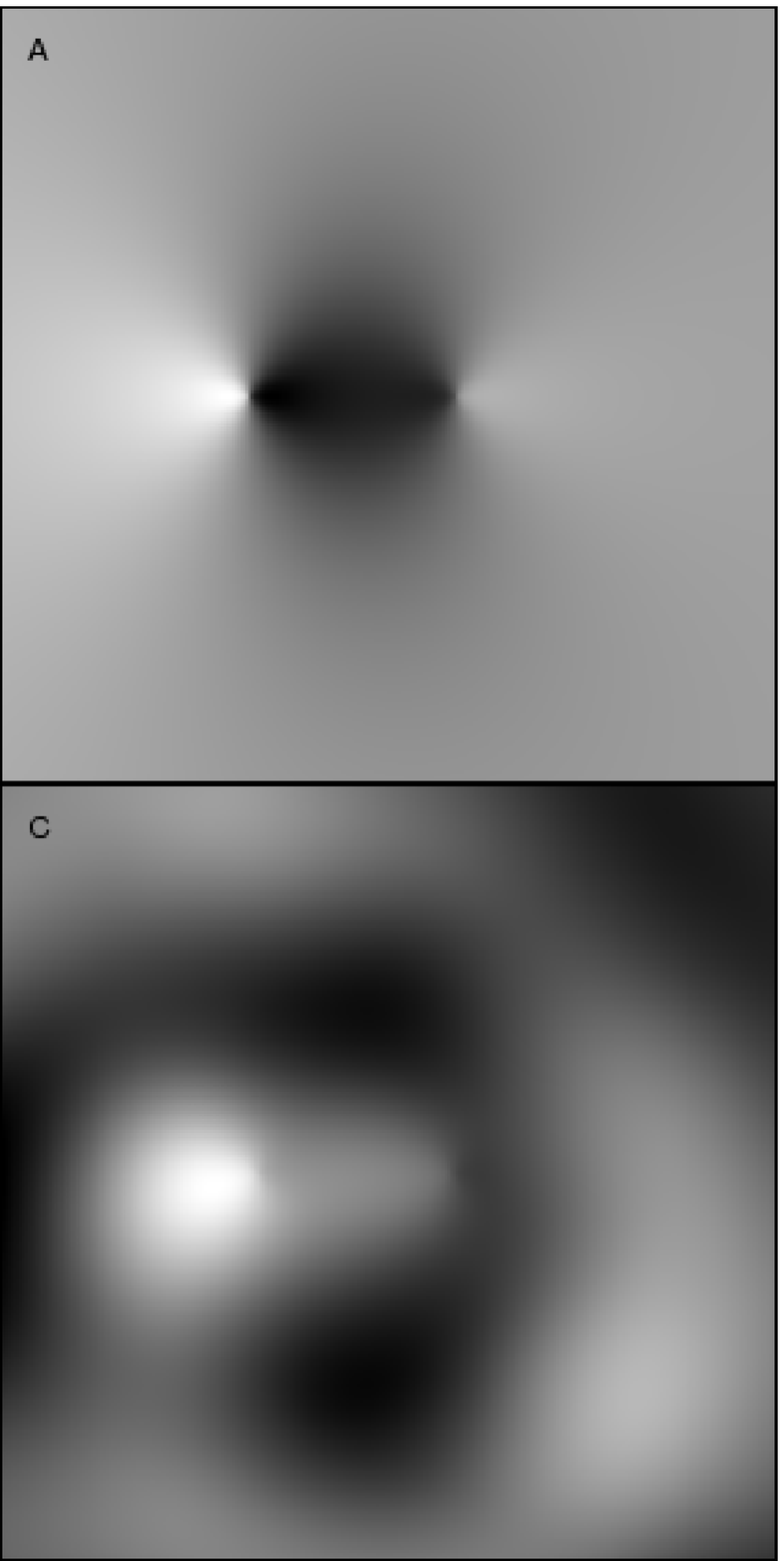}}
      \hspace{-3.5mm}
  \subfigure
      {\includegraphics[width=4.2truecm]{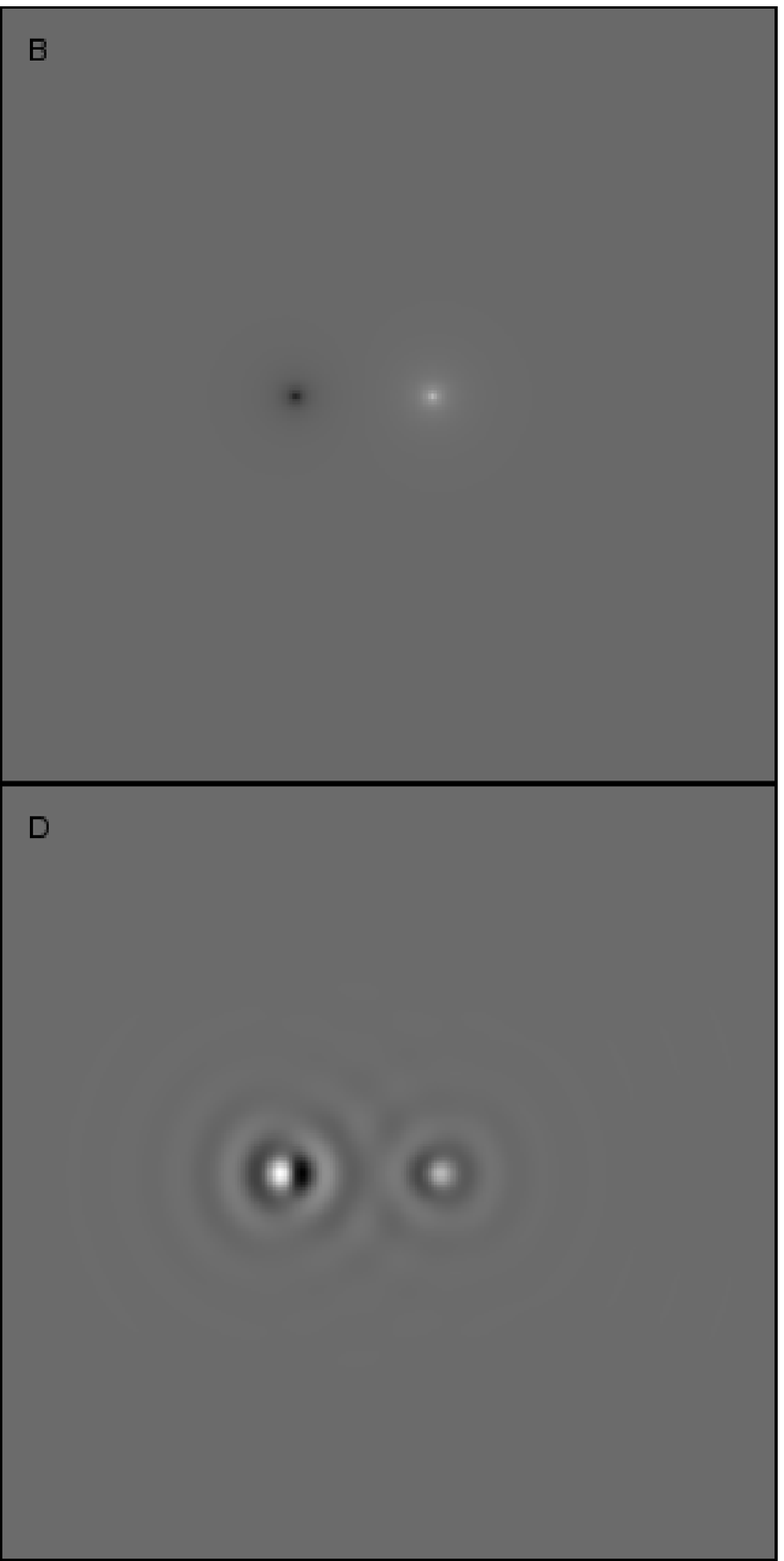}}
  \caption{ Spatial distribution of the signals considered in a major merger
      at $z=0.3$, of two haloes with masses $10^{15}M_\odot$ and $5\times
      10^{14}M_\odot$, and with an infall velocity of $1500 km/s$ at $45$
      degrees with respect to the line of sight. Since the two haloes are
      colliding, the RS effect (top left) of both components adds
      constructively, amplifying the resulting signal in the region between
      them. The kSZ effect (top right) shows a typical dipolar pattern
      reflecting opposite velocities along the line of sight. The anisotropy
      introduced by the lensing effect has a complex shape, since it depends
      on the configuration of both haloes {\em and} the CMB gradient (bottom
      left). The bottom right panel shows our filter. It is high--pass in
      order to suppress the CMB primary anisotropy noise contribution. The
      temperature amplitudes of these panels along the line containing the
      haloes are displayed in Figure~(\ref{fig:comapreF_p}). }
\end{figure}

\begin{figure*}[t]
 \centering
 \subfigure
   {\includegraphics[width=8cm]{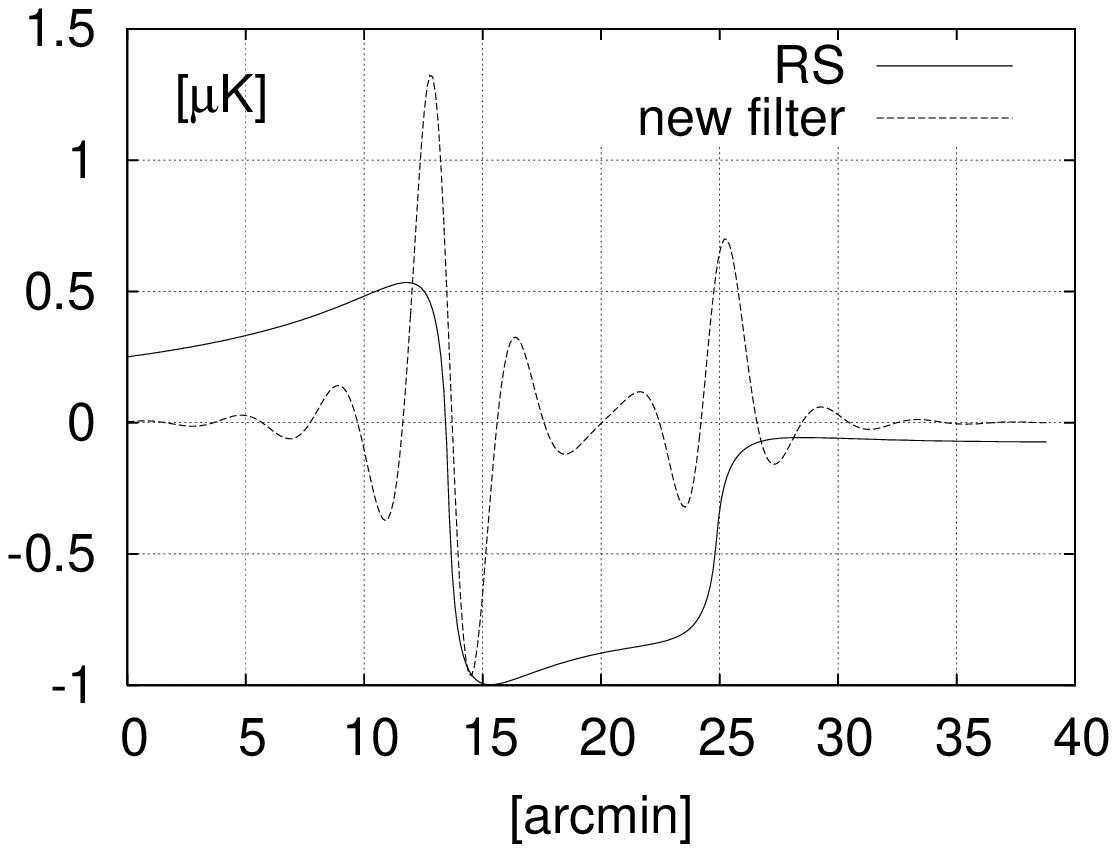}}
 \hspace{-8mm}
 \subfigure
   {\includegraphics[width=8cm]{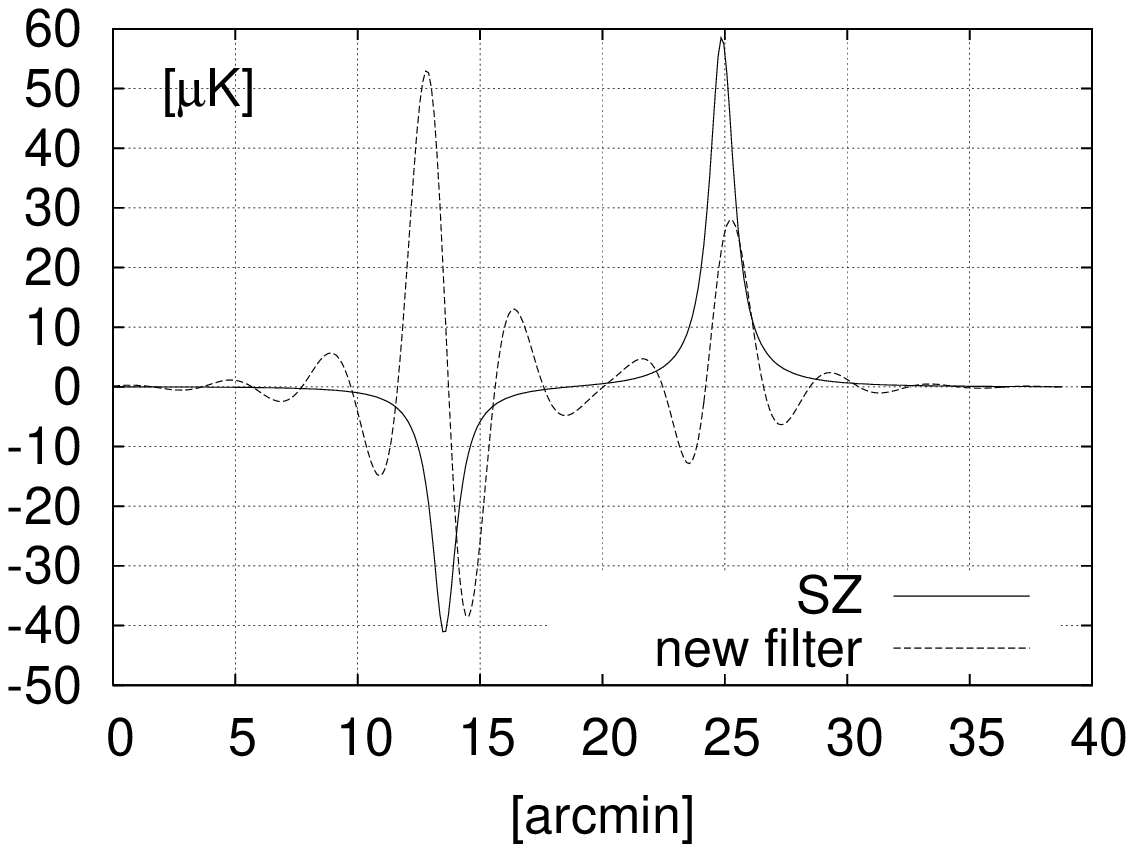}}
 \subfigure
   {\includegraphics[width=8cm]{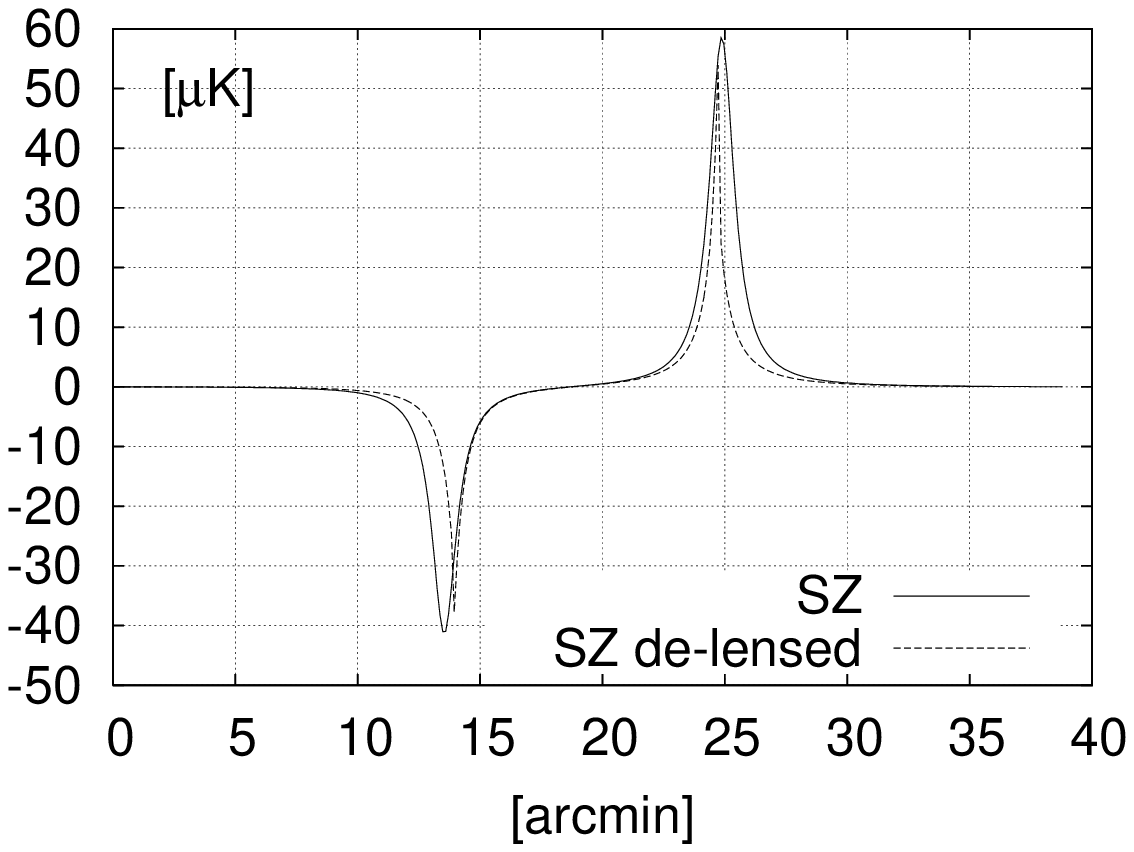}}
 \hspace{-8mm}
 \subfigure
   {\includegraphics[width=8cm]{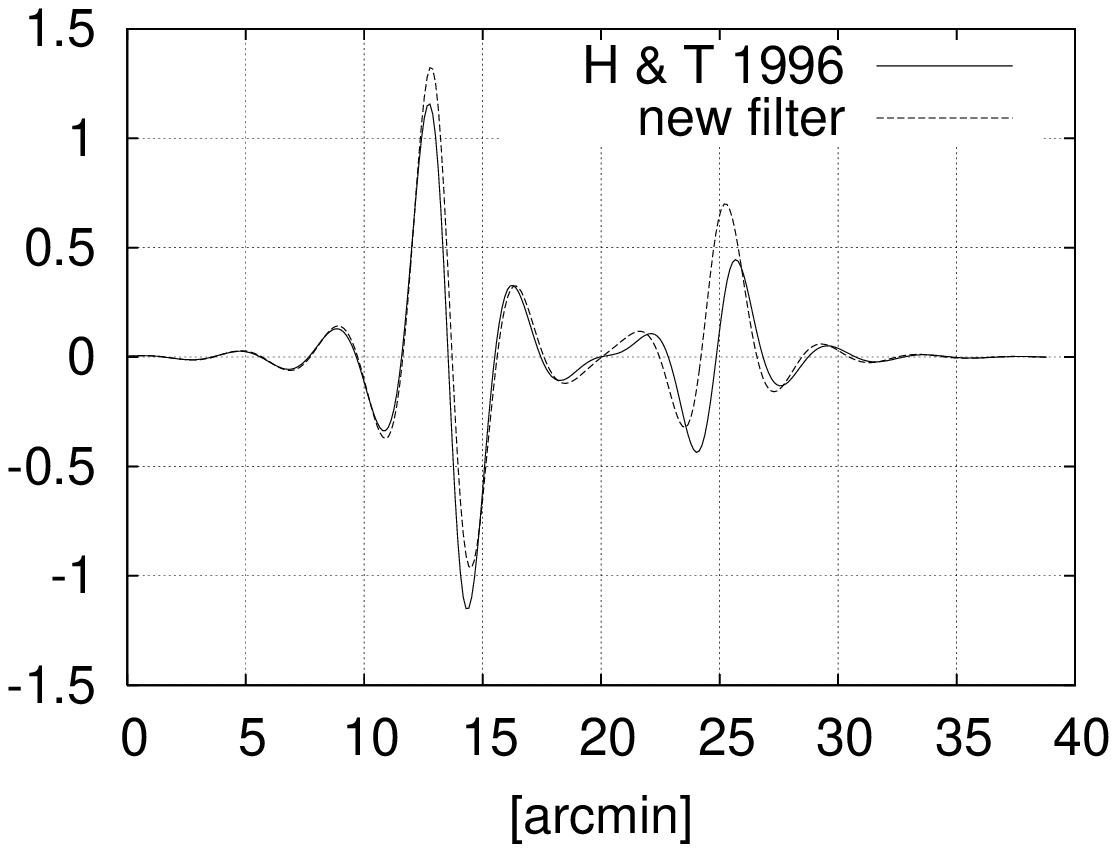}}
  \caption{Comparison between the filter profile and the different components
  under consideration. The top left panel shows the de-lensed RS effect
  superimposed on the filter. Note the sudden change of amplitude and sign of
  the filter close to the cluster cores. The top right panel displays the kSZ
  effect and the filter, whereas the bottom left one shows how the de--lensing
  procedure acts on the kSZ effect, shrinking its profile and shifting the
  peaks closer to the center. In the bottom left panel we show how the
  de--lensing procedure acts on the kSZ effect, by shrinking its profile and
  shifting the peaks closer to the center. The bottom right panel compares our
  filter (new filter) with the filter originally proposed by \cite{SE00.1} (H
  \& T 1996). The difference between these two filters is negligible when the
  contaminant given by kSZ effect (the $Z$ template) is orthogonal to the
  signal, otherwise it becomes relevant.}
\label{fig:comapreF_p}
\end{figure*}

The template for the RS effect ($\tau (\vr)$) was modeled by two merging NFW
haloes, as explained in Section~(\ref{sec:RS}). The contribution from small
merging substructures should be negligible, since our filter is a weighted
integral over the whole field (Equation~\ref{eqn:A}), and these minor mergers
are odd functions of typical scales much smaller than the one of the filter.

The kSZ masks are applied as a weight templates and are not directly
subtracted from the input data. This makes the accuracy to which the templates
are known uncritical. For this reason it is possible to use the observations
at 217 GHz (the frequency at which the tSZ vanishes to a good approximation)
as a good template for the kSZ. There might be some tSZ residuals, together
with the lensing and RS signals, but their contribution is negligible for the
filter construction. Again, this is because the filter does not subtract this
template directly from the maps, which of course would erase all the RS signal
we are searching for, but it only weights down optimally the regions
contaminated by the kSZ effect.

\section{Simulation of the observations}\label{sec:simulation}

To test the described procedure we performed simulations of the secondary
anisotropies induced on the CMB by binary cluster mergers in a cosmological
context. For the two merging haloes, we adopt a NFW profile for the dark
matter component and a spherical isothermal sphere (SIS) for the gas
distribution. We ignore tidal effects and the hydrodynamic processes involved
in the interaction between the intra-cluster gas (ICM) of the two haloes. For
our purpose, the model adopted gives a sufficiently realistic representation
of RS, kSZ, lensing effect and instrumental noise.

We adopt a standard $\Lambda$CDM cosmology with a density contribution from
dark matter, baryons and cosmological constant of $\Omega_\mathrm{DM}=0.27$,
$\Omega_\mathrm{B}=0.044$, and $\Omega_\Lambda=0.7$, respectively. The Hubble
constant was set to $H_0=100\,h\,\mathrm{km\,s^{-1}\,Mpc^{-1}}$ with $h=0.7$
\citep{BE03.3}.

We simulated the CMB primary anisotropies as Gaussian random fields with a
resolution of $1024^2$ pixels and a field of view of $120'$. The CMB power
spectrum used was computed using CMBEASY \citep{DO03.3}, assuming a
re--ionization fraction of $0.1$ at redshift $z=6.2$ and a present helium
abundance of $0.24$.  The field size ensures an adequate sampling of the CMB
multipoles, and permits the retention of only the central quarter of the whole
field, avoiding the boundary effects of the Fourier transforms.

For an isothermal cluster with negligible internal motions the kSZ and the tSZ
effects are proportional to one another:
\[ \frac{\delta T_{tSZ}}{T_0} (\theta) = g_{tSZ} (\nu ) \;\frac{k_BT_e}{m_ec^2}
                        \;\int \;dl \;n_e \sigma_T ,
\]
\begin{equation}
\frac{\delta T_{kSZ}}{T_0} (\theta ) = \frac{{-\bf v}\cdot {\bf n}}{c}
                        \; \int\; dl \;n_e \sigma_T.
\label{eq:bothsz}
\end{equation}
in the previous equation $g_{tSZ} (\nu )$ provides the (non--relativistic)
frequency dependence of the tSZ effect in temperature, and $m_e$, $n_e$ and
$T_e$ are the electron mass, number density and temperature, respectively. The
integrals are performed along the line of sight given by the pointing
direction ${\bf n}$, and the angle $\theta$ separates the line of sight from
the cluster center. The dependence of the integral on the line of sight
translates into the angular dependence. In particular, we assumed the ICM of
each merger component to be distributed as an isothermal sphere with a radial
$\beta$--profile with $\beta=1$ (i.e.~a King profile).  We based our strategy
on observations at $ 217~GHz$, where the tSZ effect vanishes. Thus the SZ
contamination is given only by the kSZ component.

The lensing effect on the CMB is described by the lens equation:
\be \label{eq:1}
  T(\vec\theta)=\tilde T[\vec\theta-\vec\alpha(\vec\theta)]\approx
  \tilde T(\vec\theta)-
  \vec\nabla\tilde T(\vec\theta)\cdot\vec\alpha(\vec\theta) \,,
\ee
where $T$ is the observed temperature map, $\tilde{T}$ is the un-lensed CMB
and $\vec{\alpha}$ is the cluster deflection angle. \cite{BA96.1} wrote an
analytic expression for the deflection angle caused by a NFW halo, and, by
using it we can compute the $\vec\nabla\tilde T\cdot\vec\alpha$ term, which
represents the secondary anisotropy given by the lensing effect
\citep{SE00.1}. The linear expansion is justified as long as the variations
of the local CMB temperature gradient are small on a deflection angle scale.

The RS and kSZ effects were computed after introducing this deflection angle
into Equation~(\ref{eqn:RS2}). The distribution of cluster mergers in the
universe is described according to the equations presented in
Section~(\ref{sec:carlos}).  The number of clusters $N_{M,z}$ was computed
with that formalism on a finite number of mass-redshift cells. Then, for each
cell, we randomly distributed $N_{M,z}$ clusters with redshifts and masses
enclosed in the cell range.

We included in our simulation the noise from a single dish telescope with a
resolution of $1$ arcminute and a sensitivity of $1 \mu K$ per beam. This
noise was modeled as a Gaussian field with the following power spectrum
\begin{equation}
  C_l^\mathrm{noise}=w^{-1}\exp\left[
  \frac{l(l+1)\,\mathrm{FWHM}^2}{8\ln2}
  \right]\;,
\end{equation}
where $\Delta T/T$ is the sensitivity of the experiment in on the scale of
the beam and
$w^{-1}:=(\Delta T/T\,\mathrm{FWHM})^2$ \citep{KN95.1}. The exponential
expresses the increasing weight of noise when probing sub-beam scales. These
characteristics of the simulated instrument are compatible with the
next--generation millimetric observatories, as shown in
Table~(\ref{tab:instruments}).  We have not included the parameters
corresponding to ALMA because our filter should have then included the
synthesized beam of the interferometer, but its sensitivity should enable it
to detect signals of amplitudes similar and smaller than those of the RS
effect considered here.

\section{Results}

In order to probe the ability of our extended filter in separating the RS
signal, we have prepared three different mock catalogues of cluster mergers
following the recipe described in Section~(\ref{sec:strategy}). These
catalogues were generated using three minimum mass thresholds of $5\times
10^{13} M_\odot$, $8\times 10^{13} M_\odot$ and $10^{14} M_\odot$, and contain
$6807$, $266$ and $28$ cluster mergers, respectively.

The analysis of the mock catalogues was carried out in two ways, corresponding
to the degree of knowledge of the template maps.

\begin{itemize} 
\item {\emph Case A}. We assume that we know the masses of the merging
   clusters with an uncertainty of $30 \%$. These values are used to build the
   kSZ and the RS templates. For the halo radial velocity, we use
   Equation~(\ref{eqn:B_i}) with an expected velocity dispersion
   $\sigma_v\approx 300 km/s$.
\item {\emph Case B}. Similar to the previous case but, instead of building a
   kSZ template, we shall use directly the millimetric observations of the kSZ
   effect, which account for the kSZ asymmetries caused by internal motions
   \citep{NA03.1}. The presence of the RS effect in these maps is negligible
   and does not affect the filter definition. The only problem related to this
   procedure is the noise present in the kSZ maps which is included in the
   filter construction. A low pass filter of the map should reduce the
   problem. In our simulation we directly used the kSZ plus the RS maps,
   assuming that a noise suppression procedure had already been applied to
   those maps.
\end{itemize}

The results for Case A are presented in Table~(\ref{tab:res1}). The columns
represent: first, the lower limit adopted to build the cluster samples;
second, the number of objects in the sample; third, the average RS signal of
the sample; and last, the estimate.

Even with a moderate size cluster merger catalogue ($266$ clusters), the
significance of our method approaches the 2-$\sigma$ level in the detection of
the RS signal.  Since we are observing different mergers, and the errors in
different mergers are uncorrelated, our Poissonian error bars scale is the
inverse square root of the number of mergers.
\begin{table}
  \caption{RS amplitude estimates for three mass selected cluster subsamples,
  defining the filter with a $30 \%$ mass uncertainty and using only the
  expected velocity dispersion $\sigma_v\approx 300~km/s$. Error bars in the
  last two columns correspond to the combined measurement from all clusters in
  the corresponding sample and with the coresponding filters (case A and B). }
  \label{tab:res1}
  \begin{center}
    \begin{tabular}{|c|c|c|c|c|}
      \hline
      mass & clusters  & simulation & estimate A & estimate B\\
      $M_\odot$ &           &  $\mu K$   & $\mu K$  & $\mu K$  \\
      \hline
      $M>5 \times 10^{13}$ &  6807 & 0.18 & 0.13 $\pm$ 0.03 & 0.16 $\pm$ 0.03\\
      $M>8 \times 10^{13}$ &  266  & 0.25 & 0.24 $\pm$ 0.14 & 0.13 $\pm$ 0.19\\
      $M>1 \times 10^{14}$ &  28   & 0.28 & 0.29 $\pm$ 0.42 & 0.05 $\pm$ 0.68\\
      \hline
    \end{tabular}
  \end{center}
\end{table}

The results for Case B are also presented in Table~(\ref{tab:res1}). As one
can see, RS signal detection is still possible. Comparing these results with
those of the previous table, it is clear that the filter efficiency is not
significantly affected. This is mainly due to the {\it statistical} nature of
our filter: it relies not on the accuracy of our templates for each merger,
but on the accuracy of our templates when describing the {\em average} spatial
properties of the kSZ and RS signals.  Indeed, in principle one could also use
the same average RS and kSZ templates for all mergers. The precision of our
method is hence limited by the number of available mergers according to
Poisson statistics.

We recall that these results have been obtained ignoring the component of the
RS signal of galaxy clusters given by the collapse of the environmental
dark-matter in the potential well of the system. The inclusion of this
component would increase the signal-to-noise ratio of the estimates.

\section{Conclusions}

We presented a method to extract the kinematic properties of major merger
events of galaxy clusters by means of their RS signal, which is observable at
centimeter/millimeter wavelengths.
The RS effect is a secondary anisotropy of the CMB produced by the time
variation of the gravitational potential along the line of sight. In this
particular case, the main contribution comes from the infall motion of the
merging components, so that the total signal is a measurement of the
convergence of the projected perpendicular momentum of the system (see Paper
I).

We proposed a method to extract this RS signal from a sample of cluster
mergers observed at millimetric wavelengths. These observations will be
contaminated by the primordial CMB fluctuations, the kinematic SZ signal and
the lensing effect from the clusters, together with the instrumental noise.
We assumed that the thermal component of the SZ effect can be subtracted by
multi-frequency observations, given that we know its frequency dependence.

The RS effect amplitude in one single merger event is much smaller than all
the other components and thus only a statistical detection of the signal may
be achieved. To maximize the signal-to-noise ratio and decrease the number of
mergers that we need to co-add, we can apply a filter which enhances the RS
signal above the other components.
Therefore, we proposed in Section~(\ref{sec:filter}) an extended version of a
matched filter presented by \cite{HN96.1}. A complete derivation of our filter
is given in Appendix~(\ref{app:filter}).  It requires as input the power
spectra of the CMB and the instrumental noise, plus two templates, one for the
RS signal and another for the contaminating kSZ effect. These templates can be
simple analytic models parameterized with the mass estimates provided by the
tSZ measurements, and with the expected average velocities. We would like to
note that the filter construction is general so that it can be applied to
other problems like the extraction of gravitational lensing on CMB maps
\citep[e.g.][]{HN96.1,MAT04.1} or on background galaxies \citep{MAT04.2}.

We showed that the developed filter can control the contamination from all
signals except CMB lensing. This is because the RS and the lensing effect
originate from the same gravitational deflection field, and thus their spatial
frequencies are similar. Therefore, we included in our pipeline an additional
step before the filter application, i.e. a ``de--lensing'' procedure. We
propose remapping the input data according to the inverse template deflection
field, again parameterized by the tSZ mass estimate.

The method was applied to mock catalogues of cluster mergers. Our results show
that the RS signal of merging clusters could be measured with this method
after observing of the order of $1,000$ cluster mergers. These numbers will be
achieved by combining the expected yields of the upcoming high--resolution
millimetric surveys (e.g. ACT, SPT, ALMA). Assuming a fraction of 30\% for
the number of mergers over the number of clusters, a RS detection will be
feasible if future SZ surveys are able to perform sensitive observations of
the order of $10^4$ clusters.

\appendix

\section{Optimal filter}\label{app:filter}

We aim to build an optimized filter $\psi(\vr)$ to extract from a noisy data
set $s(\vr)$ the best estimate of amplitude $A$ of any signal with a known
spatial shape $\tau(\vr)$:
\be \label{eqn:signal}
s(\vr)=A\tau(\vr) + n(\vr) + z(\vr) \,.
\ee
We assume $n(\vr)$ to be a Gaussian random noise with zero mean ($\langle
n(\vr) \rangle=0$) and known power spectrum $P_N(\vk)$; and
$z(\vr)=\sum_{i=1}^M v_i z_i(\vr)$ to be the sum of $M$ different noise
components with unknown amplitude $v_i$ with zero mean ($\langle v_i
\rangle=0$) but known variance $\sigma_{v \,i}$.

We define the integral of a general function $f(\vk)$ as
\be \label{eqn:integrand}
\{ f \} \equiv \int \frac{d^2k}{(2\pi)^2} f(\vk) \,.
\ee

Since the Equation~(\ref{eqn:signal}) is linear in $A$, its most general linear
estimate in the Fourier domain can be written as
\be \label{eqn:estimate}
\bar{A} = \left\{S \Psi^* \right\},
\ee
where the capital letters refer to Fourier transforms. This estimate is
required to have a bias $b=\langle\bar{A}-A\rangle=0$. It has a variance
$\sigma^2=\langle (\bar{A}-A)^2 \rangle$ given by \be \label{eqn:bias} b = A
\left( \left\{ T \Psi^* \right\} -1\right) \,, \ee
\be \label{eqn:variance}
\sigma^2 = b^2 +
           \left\{|\Psi|^2 P_n \right\} +
           \left\{ \vZ^t \Psi^* \right\} \vV \left\{ \vZ^* \Psi \right\} \,,
\ee
where we grouped the collection of the $M$ noise contributions $Z_i(\vk)$ in a
vector $\vZ(\vk)$ and defined the covariance matrix $\vV$ of its components
as $V_{ij}=\langle v_iv_j \rangle$.

Since we are interested deriving a filter which minimizes the estimate
variance maintaining the unbias condition $b=0$, we introduce the Lagrangian
multiplier $\lambda$ and search for the filter function $\Psi$ which minimizes
the action $L=\sigma^2+\lambda b$, obtaining
\be\label{eqn:filter}
\Psi(\vk) = \frac{\lambda T(\vk) - \vZ^t(\vk) \vV \left\{ \vZ^* \Psi \right\}}
                 {P_N(k)} \,,
\ee
where the apex $t$ is for the transposed vectors.

The normalization factor $\lambda$ is obtained by substituting
Equation~(\ref{eqn:filter}) in Equation~(\ref{eqn:bias}), yielding
\be \label{eqn:lambda}
\lambda = \frac{1+ \Re\left(\left\{T^*\vZ^t/P_n\right\}
                            \vV\left\{\vZ^*\Psi \right\} \right)}
               {\left\{|T|^2/P_n\right\}} \,,
\ee
where $\Re$ is for the real part.

As is clear from Equation~(\ref{eqn:filter}), the filter has to be computed
through an iterative procedure, where the starting point can be the filter
with the variance matrix $\vV$ equal to the null matrix (this is essentially
the filter originally proposed by \cite{HN96.1}). This iterative procedure is
required since the filter evaluation requires the correlation between the
filter itself and the noise sources.

The simple intuitive interpretation of this filter is that the first term
$T/P_n$ maximizes the sensitivity on the spatial frequencies where the signal
$T$ is large and the noise power spectrum $P_n$ is small.  At the same time
the second term $(\vZ^t\vV\{\vZ^*\Psi\})/P_n$ introduces a weighted mask which
takes into account the correlation between the filter $\Psi$ and all the $Z_i$
noise components, as well as the correlation between the different $Z_i$ noise
components. It is important to note that the filter $\Psi$ does not apply a
mask only by subtracting the $Z_i$ templates from the input data, leading to
spurious residuals, but it down-weights the influence of the regions where the $Z_i$ components
are dominant and correlated with the filter itself. This means it is not
critical to have a very accurate template for the noise sources, in order to
apply the filter properly. It is sufficient to have a simple model which
represents their average shapes.

The derived filter satisfies a generalized problem, where the noise sources
$v_iZ_i$ can also be correlated, but in the majority of cases the noise
sources will be uncorrelated, either due to them having different positions or
their amplitudes $\vec{v}$ being mutually uncorrelated. In this case the
matrix $\vV$ is diagonal and the filter function simplifies to
\be
\Psi(\vk) = \frac{\lambda T(\vk) - \sum_i^M B_i Z_i(\vk)}
                 {P_n(k)} \,,
\ee
where the normalization can be written as $\lambda=(1+D)/C$, the
denominator $C$ is given by
\be
C = \left\{|T|^2/P_n\right\}
\ee
and the $B_i$'s and D are defined as
\be
B_i = \sigma_{vi}^2 \{ Z_i^* \Psi \} \,,
\ee
and
\be
D = \Re \left(\left\{\frac{T^* \sum_i^M B_i Z_i}{P_n}\right\}\right) \,.
\ee
With this assumption, valid for many applications, the filter acquires this
simple expression.

\acknowledgements{This work was supported in part by an EARA fellowship spent
  in the Max-Planck-Institut f\"ur Astrophysik and the COFIN 2001 fellowship
  provided by the Bologna University.  C.H.M. and J.A.R.M acknowledge the
  financial support from the European Community through the Human Potential
  Programme under contract HPRN-CT-2002-00124 (CMBNET). C.H.M. is currently
  supported by NASA grants ADP03-0000-0092 and ADP04-0000-0093. M.M. thanks
  Giuseppe Tormen for providing the simulation used in
  Figure~(\ref{fig:RS_numeric2}).}

\end{document}